\documentclass[prd,floatfix,showpacs]{revtex4-1}
\usepackage{amsmath}
\usepackage{amsfonts}
\usepackage{amssymb}
\usepackage{graphicx}
\usepackage[margin=3.0cm]{geometry}
\usepackage{hyperref}

\newcommand\beq{\begin{equation}}      
\newcommand\beqnn{\begin{eqnarray*}}   
\newcommand\beqa{\begin{eqnarray}}     
\newcommand\beqann{\begin{eqnarray*}}  

\newcommand\eeq{\end{equation}}        
\newcommand\eeqnn{\end{eqnarray*}}     
\newcommand\eeqa{\end{eqnarray}}       
\newcommand\eeqann{\end{eqnarray*}}    

\newcommand\bi{\begin{itemize}}
\newcommand\ei{\end{itemize}}

\begin{document}

\title{Non-commutative Quantum Mechanics in Three Dimensions and Rotational Symmetry}
\date{\today}
\author{Debabrata Sinha$^{a}$, Biswajit Chakraborty$^{a,c}$ and \footnote{Corresponding author: fgs@sun.ac.za}Frederik G Scholtz$^{b,c}$}
\affiliation{$^a$S.~N.~Bose National Centre for Basic Sciences,JD Block, Sector III, Salt Lake, Kolkata-700098, India\\$^b$National Institute for Theoretical Physics (NITheP),
Stellenbosch 7600, South Africa\\
$^c$Institute of Theoretical Physics,
University of Stellenbosch, Stellenbosch 7600, South Africa}

\begin{abstract}
\noindent
We generalize the formulation of non-commutative quantum mechanics to three dimensional non-commutative space.  Particular attention is paid to the identification of the quantum Hilbert space in which the physical states of the system are to be represented, the construction of the representation of the rotation group on this space, the deformation of the Leibnitz rule accompanying this representation and the implied necessity of deforming the co-product to restore the rotation symmetry automorphism.  This also implies the breaking of rotational invariance on the level of the Schroedinger action and equation as well as the Hamiltonian, even for rotational invariant potentials.  For rotational invariant potentials the symmetry breaking results purely from the deformation in the sense that the commutator of the Hamiltonian and angular momentum is proportional to the deformation.

\end{abstract}

\maketitle

\section{Introduction}
\label{intro}
In their seminal paper, Doplicher et al.\cite{Doplicher} argued from the considerations of both general relativity and quantum mechanics that the localization of an event in spacetime with arbitrary accuracy is operationally impossible. This feature is captured by postulating non-vanishing commutation relations between operator-valued coordinates. In its simplest form they  
are given as
\begin{eqnarray}
[t,\hat{X}_{i}]=0 ;[\hat{X}_{i},\hat{X}_{j}]=i\theta_{ij},
\label{non}
\end{eqnarray}
where the time $t$ has been taken to be an ordinary c-number. This form of noncommutativity also follows from the low energy limit of string theory \cite{Seiberg}. Reformulation of Quantum mechanics or Quantum field theory based on these non-commutative relations are therefore expected to describe physics at a much higher energy scale than the conventional local Quantum Field theory and perhaps can provide another window into the nature of Planck-scale physics and complement the insights gained through other approaches like String theory and loop quantum gravity. Aside from the high energy considerations, this kind of non-commutative structure also has relevance in condensed matter physics like the Quantum Hall effect \cite{Hall} and topological insulators \cite{Prodan}.

 The general point of view regarding the matrix $\Theta = \lbrace \theta_{ij}\rbrace$ is that the entries are constant, as if they are new constants of nature like $\hbar$, $c$, $G$ etc \cite{Asch} and $\theta_{ij}$ do not transform as a second-rank tensor under $SO(3)$. One therefore does not expect the coordinates $\hat{X}_{i}$ to transform vectorially rendering the construction of a $SO(3)$ invariant scalar potential virtually impossible. This problem, however, does not arise in $D=2$, as $\Theta$ remains invariant under $SO(2)$ rotations in this case, even if $\theta_{ij}$ is subjected to a tensorial transformation and one can easily construct a $SO(2)$ invariant potential. Indeed, in \cite{Scholtz} an analytical solution to the problem of a particle, confined in a $2D$ spherical infinite potential well, was provided in a completely operatorial approach, bypassing the conventional approach of using the Moyal/Voros star product. Since these star products are naturally associated with respective bases \cite{Basu}, the analysis in \cite{Scholtz} is completely independent of any choice of basis and has a general validity.
 
 It should, however, be pointed out that the $2D$ case is rather trivial and non-trivialities arise only in $D\geq 3$. It is therefore desirable to understand whether it is possible to construct a $SO(3)$ invariant potential in $3D$ in a completely operatorial approach in the spirit of \cite{Scholtz}.
 
 An attempt in this direction was made in \cite{Chakra} in a Hopf algebraic approach, where the deformed co-product was used to define a deformed adjoint action and the associated deformed brackets. Interestingly, it was observed in \cite{Chakra} that the non-commutative coordinates transform covariantly under these deformed brackets when the angular momentum operator is also deformed simultaneously. However, even this approach failed to produce a $SO(3)$ invariant potential with respect to these deformed brackets, even if one starts with a $SO(3)$ invariant potential in the commutative case; they are found to be afflicted with anomalies.
 
 This therefore motivates us to first generalize the operator method introduced in \cite{Scholtz} and also in \cite{Gouba}, where the interpretational aspects were studied, to $3D$. This then paves the way to an understanding of the way the symmetry manifests itself on the level of the action,  Hamiltonian and Schroedinger equation, which, as far as we can establish, has not been done systematically in the literature.
 
The paper is organized as follows: In section \ref{2d} we briefly review the 2 dimensional construction to fix conventions and notations.  In section \ref{3d} the 3 dimensional generalization is introduced.  In section \ref{moyvor} we discuss the Moyal and Voros basis representations of the abstract construction introduced in \ref{3d} and make contact with the more standard non-commutative formulation in terms of Moyal and Voros star products.  Section \ref{angmom} constructs the representation of the rotation group on the quantum Hilbert space introduced in section \ref{3d}.  Section \ref{autosec} discusses the deformed co-product required to restore the rotational symmetry automorphism and section \ref{theta} shows that under this deformation the non-commutative matrix is indeed invariant. In section \ref{consangmom} the breaking of rotational symmetry on the level of the Schroedinger equation, even for rotational invariant potentials, is discussed.  Finally, section \ref{conc} summarizes and draws conclusions. 

\section{Review of two dimensional non commutative quantum mechanics}
\label{2d}
The non-commutative Heisenberg algebra in two dimension can be written as (we work in unit $\hbar=1$)
\begin{eqnarray}
[\hat{x_{i}},\hat{x_{j}}]= i\theta_{ij},
\end{eqnarray}
\begin{eqnarray}
[\hat{x_{i}},\hat{p_{j}}]= i\delta_{ij},
\end{eqnarray}
\begin{eqnarray}
[\hat{p_{i}},\hat{p_{j}}]=0.
\end{eqnarray}
One can construct standard creation and annihilation operators $b^{\dagger}$ and $b$:\\
\begin{eqnarray}
\label{createann}
b=\frac{\hat{x_{1}}+i\hat{x_{2}}}{\sqrt{2\theta}}, b^{\dagger}=\frac{\hat{x_{1}}-i\hat{x_{2}}}{\sqrt{2\theta}}.
\end{eqnarray}
The non-commutative plane can therefore be viewed as a boson Fock space spanned by the eigenstate $|n\rangle$ of the operator $b^{\dagger}b$. We refer to it as the classical configuration space $\mathcal H_{c}$:
\begin{eqnarray}
\mathcal H_{c}= span\lbrace|n\rangle =\frac{1}{\sqrt{n!}}(b^{\dagger})^{n}|0\rangle \rbrace.
\end{eqnarray}
Note that this space plays the same role as the classical configuration space $\mathcal R^{2}$ in commutative quantum mechanics. Next we introduce the quantum Hilbert space in which the states of the system and the non-commutative Heisenberg algebra are to be represented. This is taken to be the set of all bounded trace-class operators (the Hilbert-Schmidt operators) over $\mathcal H_{c}$ and we refer to it as the quantum Hilbert space,$\mathcal H_{q}$,
\begin{eqnarray}
\mathcal H_{q}=\lbrace\psi:tr_{c}(\psi^{\dagger}\psi)<\infty\rbrace.
\end{eqnarray}
Note that we can also think of the states in the quantum Hilbert space as functions $\psi\left(\hat{x}_1,\hat{x}_2\right)$ (with an appropriate ordering prescription) of the non-commuting coordinates $\hat{x}_1$, $\hat{x}_2$ as they are essentially the trace class operators generated by the bounded Weyl operators associated with $\hat{x}_1$ and $\hat{x}_2$. Physical states are represented by the elements of $\mathcal H_{q}$ and are denoted by a round bracket $ \psi\equiv |\psi)$. The inner product is defined as
\begin{eqnarray}
\label{qhinner}
(\phi |\psi)=tr_{c}(\phi^{\dagger}\psi),
\end{eqnarray}
where the subscript $c$ refers to tracing over $\mathcal H_{c}$. We reserve $\dagger$ to denote hermitian conjugation on the classical Hilbert space, while $\ddagger$ denotes hermitian conjugation on the quantum Hilbert space.  If $\hat{X_{i}}$,$\hat{P_{i}}$ are the representations of the operators $\hat{x_{i}}$ and $\hat{p_{i}}$ acting on $\mathcal H_{q}$, then a unitary representation, i.e. $\hat X_i^\ddagger=\hat X_i$ and $\hat P_i^\ddagger=\hat P_i$, is obtained by the following action:
\begin{eqnarray}
\hat{X_{i}}\psi= \hat{x_{i}}\psi,\\
\hat{P_{i}}\psi=\frac{1}{\theta}\epsilon_{ij}[\hat{x_{j}},\psi].
\end{eqnarray}
It is easily verified that the momentum eigenstate $|p)$ are given by
\begin{eqnarray}
|p)=\sqrt{\frac{\theta}{2\pi}}e^{ip.\hat{x}}, \hat{P_{i}}|p)=p_{i}|p)
\label{mom1}
\end{eqnarray}
and that they satisfy the usual resolution of identity and orthogonality condition
\begin{eqnarray}
\int d^{2}p|p)(p|=1_q,(p|p')=\delta^{2}(p-p').
\label{mom2}
\end{eqnarray}
Following the analogy of coherent states of the Harmonic oscillator, one can introduce minimum uncertainty states in the classical configuration space
\begin{eqnarray}
|z\rangle = e^{-\bar{z}b+zb^{\dagger}}|0\rangle = e^{-\frac{1}{2}|z|^{2}}e^{zb^{\dagger}}|0\rangle \in \mathcal H_{c},
\end{eqnarray}
satisfying
\begin{eqnarray}
b|z\rangle =z |z\rangle
\end{eqnarray}
for an arbitrary complex number $z$. From this a basis $|z,\bar{z})=|z\rangle \langle z| \in \mathcal H_{q}$ can be constructed for the quantum Hilbert space. In particular they satisfy
\begin{eqnarray}
B|z,\bar{z})&=& z|z,\bar{z}),\\
(z',\bar{z'}|z,\bar{z})&=& tr_{c}[(|z'\rangle \langle z'|)^{\dagger}(|z\rangle \langle z|)]= e^{-|z-z'|^{2}}
\end{eqnarray}
and, most importantly, the completeness relation
\begin{eqnarray}
\int \frac{d^{2}z}{\pi}|z,\bar{z}) \star_{V} (z,\bar{z}| = 1_q.
\end{eqnarray}
Here $B=\frac{\hat{X_{1}}+i\hat{X_{2}}}{\sqrt{2\theta}}$ is the representation of the operator $b$ on $\mathcal H_{q}$ and the Voros-star product $\star_{V}$ takes the form
\begin{eqnarray}
\star_{V}= e^{\overleftarrow\partial_{z}\overrightarrow\partial_{\bar{z}}}=e^{\frac{i}{2}\Theta^V_{ij}\overleftarrow\partial_{i}\overrightarrow\partial_{j}}
\end{eqnarray}
with
\begin{eqnarray}
\Theta^V=\left(
\begin{array}{cc}
-i\theta & \theta \\
-\theta & -i\theta
\end{array}
\right).
\label{2dtvoros}
\end{eqnarray}

We refer to this basis as the Voros basis. The overlap of this basis with a momentum eigenstate is given by
\begin{eqnarray}
(z,\bar{z}|p)&=& \sqrt{\frac{\theta}{2\pi}}e^{-\frac{\theta p^{2}}{4}}e^{i\sqrt{\frac{\theta}{2}}(p\bar{z}+\bar{p}z)}\nonumber\\
&=& \sqrt{\frac{\theta}{2\pi}}e^{-\frac{\theta p^{2}}{4}}e^{ip.x},
\label{basis}
\end{eqnarray}
where we have introduced the Cartesian coordinates
\begin{eqnarray}
x_{1}= \sqrt{\frac{\theta}{2}}(z+\bar{z}), x_{2}= i\sqrt{\frac{\theta}{2}}(\bar{z}-z)
\end{eqnarray}
so that the Voros states can alternatively be labeled as $|x)_{V}\equiv |z,\bar{z})$. From these we infer that we may expand the Voros basis as follow in terms of momentum states
\begin{eqnarray}
|x)_{V}= \sqrt{\frac{\theta}{2\pi}}\int d^{2}p e^{-\frac{\theta p^{2}}{4}}e^{-ip.x}|p)=\int \frac{d^{2}p}{2\pi}\theta e^{-\frac{\theta p^{2}}{4}}e^{ip.(\hat{x}-x)}.
\label{voros}
\end{eqnarray}
Next we introduce what we refer to as the Moyal basis, defined as an expansion in terms of momentum states as follows
\begin{equation}
\label{moyalba}
|x)_M=\int \frac{d^2p}{2\pi}e^{-ip\cdot x}|p)=\sqrt{\frac{\theta}{2\pi}}\int\frac{d^2p}{2\pi}e^{ip\cdot(\hat x-x)}.
\end{equation}
These states satisfy
\begin{eqnarray}
\int d^2x |x)_M\star_M{}_M(x|&=&\int d^2x|x)_M{}_M(x|=1_q,\nonumber\\
\label{mcomplt2}
(p|x)_M&=&\frac{1}{2\pi}e^{-ip\cdot x},\nonumber\\
{}_M(x|x^\prime)_M&=&\delta^2(x-x^\prime ),
\label{mortho}
\end{eqnarray} 
with
\begin{eqnarray}
\star_{M}=e^{\frac{i}{2}\Theta^M_{ij}\overleftarrow\partial_{i}\overrightarrow\partial_{j}}
\end{eqnarray}
and
\begin{eqnarray}
\Theta^M=\left(
\begin{array}{cc}
0 & \theta \\
-\theta & 0
\end{array}
\right).
\label{2dtmoyal}
\end{eqnarray}
 
The Moyal basis is therefore an orthogonal basis, unlike the Voros basis. Using (\ref{voros}) we find the overlap between the Moyal and Voros basis vectors to be 
\begin{equation}
{}_V(x^\prime|x)_M=\sqrt{\frac{2}{\pi \theta}}e^{-\frac{(x-x^\prime)^2}{\theta}}.
\label{QMVprodct}
\end{equation} 
Clearly, in the commutative limit $\theta\rightarrow 0$ the Gaussian occurring on the RHS goes over to $2D$ Dirac delta function $\delta^{2}(x-x')$, indicating that the difference between the Moyal and Voros basis disappears in this limit.

In ${\mathcal H}_q$ one can define commuting operators $\hat X^c_i$ as
\begin{eqnarray}
\hat X^c_i=\hat X_i+\frac{\theta}{2}\epsilon_{ij}\hat P_j,
\label{XC}
\end{eqnarray}
for which the Moyal basis states are simultaneous eigenstates:
\begin{eqnarray}
\hat X^c_i|x)_M=x_i |x)_M.
\end{eqnarray}
Since ${\mathcal H}_q$ is a Hilbert space of operators, one can define a multiplication map, $m:{\mathcal H}_q\otimes {\mathcal H}_q\rightarrow {\mathcal H}_q$, on it, given by $m\left(|\psi)\times|\phi)\right)=|\psi\phi)$, which turns this space into an operator algebra.  Expanding a generic state $|\psi)$ as
\begin{equation}
|\psi)=\sqrt{\frac{\theta}{2\pi}}\int\frac{d^2p}{2\pi}\psi(p)e^{ip\cdot \hat x},
\end{equation}
(note that the condition of normalizability of the state, i.e., $(\psi|\psi)={\rm tr_c}(\psi^\dagger\psi)<\infty$ implies that the function $\psi(p)$ must be square integrable) one easily verifies the following composition rules when the product state is represented in the Moyal or Voros basis:
\begin{eqnarray}
\label{map1}
{}_M(x|\psi\phi)&=&\sqrt{2\pi\theta} {}_M(x|\psi)\star_M{}_M(x|\phi),\\
\label{map2}
{}_V(x|\psi\phi)&=&4\pi^2{}_V(x|\psi)\star_V{}_V(x|\phi).
\end{eqnarray}
Here
\begin{eqnarray}
\label{moywf}
{}_M(x|\psi)&=&\int\frac{d^2p}{(2\pi)^2}\psi(p)e^{ip\cdot x},\\
\label{vorwf}
{}_V(x|\psi)&=&\sqrt{\frac{\theta}{2\pi}}\int\frac{d^2p}{(2\pi)^2}\psi(p)e^{-\frac{\theta p^2}{4}}e^{ip\cdot x}=\sqrt{\frac{\theta}{2\pi}}e^{\frac{\theta \nabla^2}{4}}{}_M(x|\psi).
\label{voros3}
\end{eqnarray} 

\section{The three dimensional generalization}
\label{3d}
We begin with the algebra satisfied by the coordinate operators:
\begin{eqnarray}
 [\hat{x}_{i},\hat{x}_{j}]=i\theta_{ij}=i\epsilon_{ijk}\theta_{k}; i,j,k=1,2,3.
 \label{non1}
 \end{eqnarray}
Since $\theta_{ij}$ is $3\times 3$ antisymmetric matrix, it must be degenarate and one can make a suitable $SO(3)$ transformation to orient the real vector $\vec{\theta}$ ($\lbrace \theta_{k}=\frac{1}{2}\epsilon_{ijk}\theta_{ij}\rbrace$) along the $3$rd axis. To be specific, this is accomplished by making the transformation
\begin{eqnarray}
\hat{x}_{i}\rightarrow \hat{\bar{x}}_{i}={\bar R_{ij}}{\hat x}_{j}
\label{non2}
\end{eqnarray}
with
\begin{eqnarray}
\bar{R}=\left(
\begin{array}{ccc}
\cos\alpha \cos\beta & \sin\beta \cos\alpha & -\sin\alpha\\
-\sin\beta & \cos\beta & 0\\
\sin\alpha \cos\beta & \sin\alpha \sin\beta & \cos\alpha
\end{array}
\right)
\label{matrix1}
\end{eqnarray}
if the vector $\vec{\theta}$ is parametrised as
\begin{eqnarray}
\vec{\theta}=\theta \left(
\begin{array}{c}
\sin\alpha \cos\beta\\
\sin\alpha \sin\beta\\
\cos\alpha
\end{array}
\right).
\label{matrix2}
\end{eqnarray}
Note that this form of $\bar{R}$ is not unique as we still have a freedom to make an additional $SO(2)$ rotation around the $\vec{\theta}$ axis. With this the NC coordinate algebra assumes the form 
\begin{eqnarray}
[\hat{\bar{x}}_{1},\hat{\bar{x}}_{2}]=i\theta,\nonumber \\\
[\hat{\bar{x}}_{1},\hat{\bar{x}}_{3}]=0,\nonumber\\\ [\hat{\bar{x}}_{2},\hat{\bar{x}}_{3}]=0.
\label{nc}
\end{eqnarray}
In this barred frame the non-commutative matrix, $\bar\Theta_{ij}={\bar\theta}_{ij}$, therefore takes the form 
\begin{eqnarray}
\label{thetab}
\bar{\Theta}=\bar{R}\Theta \bar{R}^{T}=\left(
\begin{array}{ccc}
0 & \theta & 0\\
-\theta & 0 & 0\\
0 & 0 & 0
\end{array}
\right).
\label{matrix3}
\end{eqnarray}
We therefore see that the $\hat{\bar{x}}_{3}$ coordinate essentially becomes commutative and a particle undergoing motion in this frame finds itself moving in a space which is nothing but the direct product of the $2D$ non-commutative plane, introduced in the previous section, and the real line. From this perspective one can therefore construct the classical configuration space as a tensor product space of the non-commutative 2D classical configuration space (boson Fock space) and a one dimensional Hilbert space spanned by the eigenstates of $\hat{\bar{x}}_{3}$, i.e.,
\begin{eqnarray}
\mathcal H_{c}^{(3)}= span \lbrace{|n,\bar{x}_{3}\rangle}\rbrace= span \lbrace{|z,\bar{x}_{3}\rangle}\rbrace.
\label{hilbert}
\end{eqnarray}
Here $n$ labels the eigenstates of $b^\dagger b$, where $b^\dagger$ and $b$ were introduced in eq.(\ref{createann}), and $\bar{x}_{3}$ labels the eigenstates of $\hat{\bar{x}}_3$.  Alternatively, as described in the previous section, one can introduce a coherent state basis, labeled by $z$, for the boson Fock space. 

The action of the original (i.e. unbarred) coordinates $\hat{x}_{i}$ on these basis states is then extended through linearity, by inverting (\ref{non2}) and using the fact that $\bar{x}_3$ is the eigenvalue of $\hat{\bar{x}}_3$ as:
\begin{eqnarray}
\hat{x}_{i}|n,\bar{x}_{3}\rangle &=& [(\bar{R}^{-1})_{ij}\hat{\bar{x}}_{j}]|n,\bar{x}_{3}\rangle\nonumber\\
&=& (\bar{R}^{-1})_{i\alpha} \hat{\bar{x}}_{\alpha}|n,\bar{x}_{3}\rangle +(\bar{R}^{-1})_{i3}\bar{x}_{3}|n,\bar{x}_{3}\rangle, 
\label{bar}
\end{eqnarray}
where $\alpha,\beta =1,2$.  Note that the action of $\hat{\bar{x}}_\alpha$ is defined through the action of the creation and annihilation operators of eq. (\ref{createann}).

The next step is to define the quantum Hilbert space $\mathcal H_{q}^{(3)}$, the elements of which represent the physical states, and on which the non-commutative Heisenberg algebra is to be represented.

In analogy with ordinary quantum mechanics, where the Hilbert space of states is the space of square integrable functions of coordinates, here it becomes, as in the previous section, the Hilbert space of all functions of the operator valued coordinates (all operators generated by the Weyl algebra associated with $\hat{\bar x}_\alpha, \hat{\bar x}_3$) and the elements of the quantum Hilbert space are therefore operators acting on classical configuration space. The requirement of square integrability gets replaced by the trace class condition.  We therefore identify the appropriate quantum Hilbert space $\mathcal H_{q}^{(3)}$ to be
\begin{equation}
\label{3dhilb}
\mathcal H_{q}^{(3)}=\lbrace\psi(\hat{\bar{x}}_i):tr_c\psi^\dagger\psi<\infty\rbrace.
\end{equation}
If we use as basis for the classical configuration space eigenstates of $\hat{\bar{x}}_3$, we can replace $\hat{\bar{x}}_3$ by its eigenvalue $\bar{x}_3$ and identify the quantum Hilbert space with 
\begin{equation}
\mathcal H_{q}^{(3)}=\lbrace\psi(\hat{\bar{x}}_i):\int \frac{d\bar{x}_3}{\sqrt{\theta}}tr^\prime_c\psi^\dagger\psi<\infty\rbrace,
\end{equation}
where $tr_c^\prime$ denotes the restricted trace over the non-commutative 2D plane.  This space is therefore simply a one parameter family of quantum Hilbert spaces for the 2D non-commutative plane, which is not surprising given that the classical configuration space is simply a one parameter family of 2D non-commutative planes.  In contrast to the previous section, it is, however, important to note that this space does not coincide with the space of all Hilbert-Schmidt operators acting on classical configuration space. To emphasize that this space is the space of operators generated by the barred operator valued coordinates, we have explicitly indicated the functional dependence in (\ref{3dhilb}).  Indeed, from the discussion above it is immediately clear that elements of the quantum Hilbert space leaves the subspace $span\{|n,\bar{x}_3\rangle\}$, for fixed $\bar{x}_3$, of the classical configuration space invariant.  Hence the elements of quantum Hilbert space must constitute a subset of all the possible operators on classical configuration space.  These can be characterized as all Hilbert-Schmidt operators $\psi$ on classical configuration space that also satisfy the additional constraint $\left[\hat{\bar{x}}_3,\psi\right]=0$, i.e.:
\begin{equation}
\label{3dquan}
\mathcal H_{q}^{(3)}=\lbrace\psi:\left[\hat{\bar{x}}_3,\psi\right]=0;\;tr_c\psi^\dagger\psi<\infty\rbrace,
\end{equation}
with inner product as in (\ref{qhinner}).  This is also a more convenient characterization of the quantum Hilbert space as we now do not need to specify the operators that are used to generate the elements of the quantum Hilbert space.  We can therefore drop the functional dependence on the barred coordinates and simply denote the elements of the quantum Hilbert space by $\psi\equiv|\psi)$, keeping in mind the constraint in (\ref{3dquan}).  

To define the action of the momentum operators on the quantum Hilbert space it is convenient to introduce a further 'coordinate' $\hat{\bar{x}}_4$ such that 
\begin{equation}
\left[\hat{\bar{x}}_j,\hat{\bar{x}}_4\right]=i\theta\delta_{j3};\;j=1,2,3,
\end{equation}
i.e., it commutes with $\hat{\bar{x}}_1$ and $\hat{\bar{x}}_2$ and is conjugate to $\hat{\bar{x}}_3$. Here $\theta$ was defined in (\ref{matrix2}) and (\ref{thetab}).  Formally $\hat{\bar{x}}_4=-i\theta \frac{\partial}{\partial\bar{x}_3}$.  In terms of these the action of the momentum operators in the barred frame on the quantum Hilbert space can be expressed through the adjoint action:
\begin{equation}
\label{act}
\hat{\bar{P}}_{\alpha}\psi=\frac{1}{\theta}\Gamma_{\alpha\beta}[\hat{\bar{x}}_{\beta},\psi] ; \alpha,\beta=1,2,3,4,
\end{equation}
where 
\begin{eqnarray}
\Gamma=\left(
\begin{array}{cccc}
0 & 1 & 0 & 0\\
-1 & 0 & 0 & 0\\
0 & 0 & 0 & 1\\
0 & 0& -1&0 
\end{array}
\right).
\label{matrixGamma}
\end{eqnarray}
Note that due to the constraint on $\psi$, we have $\hat{\bar{P}}_4\psi=0$ so that there are only three non trivial momenta.

This can be extended to the action of the components of momenta in the original frame (this will referred as the fiducial frame later in the paper). Invoking linearity gives:
\begin{eqnarray}
\label{act1}
\hat{P_{i}}\psi&=&(\bar{R}^{-1})_{ij}\hat{\bar{P_{j}}}\psi\nonumber\\
&=&\frac{1}{\theta}(\bar{R}^{-1})_{ij}\Gamma_{j\alpha}[\hat{\bar{x}}_{\alpha},\psi];\;i,j=1,2,3;\;\alpha=1,2,3,4.
\end{eqnarray}

Like-wise one can introduce the position operators $\hat{X}_{i}$ acting on $\mathcal H_{q}^{(3)}$ through left multiplication, i.e., by the linear maps
\begin{eqnarray}
\label{position}
\hat{X}_{i}:|\psi) \rightarrow \hat{X}_{i}|\psi)=|\hat{x_{i}}\psi)
\end{eqnarray}
in the original fiducial unbarred frame, so that these pairs of observables now satisfy the non-commutative Heisenberg algebra
\begin{eqnarray}
[\hat{X_{i}},\hat{X_{j}}]=i\theta_{ij}; [\hat{X_{i}},\hat{P_{j}}]=i\delta_{ij}; [\hat{P_{i}},\hat{P_{j}}]=0.
\end{eqnarray}

Simultaneous eigenstates of the above commuting momentum operators will play an important role in what follows. It is a simple matter to verify that these are given by:
\begin{eqnarray}
\hat{P_{i}}|\vec{p})=p_{i}|\vec{p})
\end{eqnarray} 
where
\begin{eqnarray}
|\vec{p})&=& \frac{\theta^{3/4}}{2\pi}e^{ip_{i}\hat{x}_{i}};\;i=1,2,3\nonumber\\
&=& \frac{\theta^{3/4}}{2\pi}e^{i\bar{p}_{i}\hat{\bar{x}}_{i}}\nonumber\\
&=& \frac{\theta^{3/4}}{2\pi}e^{i\bar{p}_{\alpha}\hat{\bar{x}}_{\alpha}}e^{i\bar{p}_{3}\hat{\bar{x}}_{3}};\;\alpha=1,2.
\end{eqnarray}
Note that $\hat{\bar{P}}_4|p)=0$ as required by the constraint in (\ref{3dquan}).  Here we have also noted that 
$p_{i}\hat{x}_{i}$  is a scalar under a $SO(3)$ rotation. In complete analogy with the two dimensional case one can verify that these states satisfy the orthogonality relations 
\begin{eqnarray}
\label{3dort}
(\vec{p'}|\vec{p})&=&\frac{\theta^{3/2}}{(2\pi)^2}tr_c(e^{-i\vec{p'}.\hat{\vec{x}}}e^{i\vec{p}.\hat{\vec{x}}})\\
&=&tr^\prime_c[e^{-i\bar{p'}_{\alpha}\hat{\bar{x}}_{\alpha}}e^{i\bar{p}_{\beta}\hat{\bar{x}}_{\beta}}] \int \frac{d\bar{x}_3}{\sqrt{\theta}}[e^{-i\bar{p'}_{3}\bar{x}_{3}}e^{i\bar{p}_{3}\bar{x}_{3}}]\\
&=&\delta^{3}(\vec{p}-\vec{p'}),
\end{eqnarray}
and completeness relation
\begin{equation}
\label{3dcomp}
\int d^{3}p|p)(p|=1_q.
\end{equation}

\section{Moyal and Voros bases in three dimensions}
\label{moyvor}
The position operator $\hat{X}_{i}$  introduced in the previous section has been taken to be a left action by default. One can, likewise, construct a right action so that we can write for both left and right actions
\begin{eqnarray}
\label{op3}
\hat{X_{i}}^{(l)}\psi=\hat{x}_{i}\psi,\\
\hat{X_{i}}^{(r)}\psi=\psi\hat{x}_{i}.
\label{op4}
\end{eqnarray}
In analogy with the construction given in \cite{Pinzul}, we can also introduce here the map $\hat{X_{i}}^{(c)}$, which is the average of above left and right actions:
\begin{eqnarray}
\hat{X_{i}}^{(c)}\psi\equiv \frac{1}{2}[\hat{X_{i}}^{(l)}+\hat{X_{i}}^{(r)}]\psi.
\label{aver}
\end{eqnarray}
By splitting $\hat{x_{i}}\psi(\hat{x_{i}})$ into symmetric and anti symmetric parts, this can be rewritten as
\begin{eqnarray}
\hat{X_{i}}^{(l)}\psi=\hat{X_{i}}^{(c)}\psi+\frac{1}{2}[\hat{x}_{i},\psi].
\label{antisym}
\end{eqnarray}
The transition to the barred frame and then back to the original unbarred frame  allows us to rewrite this, using (\ref{non2}) and (\ref{act}), in the form
\begin{eqnarray}
\hat{X_{i}^{(l)}}\psi&=&\hat{X_{i}}^{(c)}\psi+\frac{1}{2}\bar{R}^{T}_{ij}[\hat{\bar{x_j}},\psi]\nonumber\\
&=& \hat{X_{i}}^{(c)}\psi-\frac{\theta}{2}\bar{R}^{T}_{ij}\Gamma_{j\alpha}(\hat{\bar{P_{\alpha}}} \psi)\nonumber\\
&=& \hat{X_{i}}^{(c)}\psi-\frac{\theta}{2}\bar{R}^{T}_{ij}\Gamma_{jk}(\hat{\bar{P_{k}}} \psi)\nonumber\\
&=& \hat{X_{i}}^{(c)}\psi-\frac{1}{2}\left(\bar{R^{T}}\bar\Theta\bar{R}\right)_{il}(P_{l}\psi).
\end{eqnarray}
where $\alpha,\beta=1,2,3,4$, $i,j,k,l=1,2,3$ and we used $\hat{\bar{P}}_4\psi=0$. Using the covariant transformation property (\ref{matrix3}) of $\Theta$ and the fact that $\psi$ is an arbitrary state, we can write this as an operator identity on $\mathcal H_{q}$: 
\begin{eqnarray}
\hat{X_{i}}^{(c)}=\hat{X}^{(l)}_{i}+\frac{\theta_{ij}}{2}\hat{P}_{j}
\label{comm}
\end{eqnarray}
satisfying
\begin{eqnarray}
[\hat{X_{i}}^{(c)},\hat{X_{j}}^{(c)}]=0.
\end{eqnarray}
The $\hat{X}_{i}$ must be regarded as the physical position operators appropriate for the quantum Hilbert space $\mathcal H^{(3)}_{q}$. $\hat{X_{i}}^{(c)}$ are the corresponding commuting position operators acting on $\mathcal H^{(3)}_{q}$, but they do not have the same physical status as the $\hat{X}_{i}$.

In analogy with (\ref{moyalba}) we introduce the normalized 'Moyal basis' as
\begin{eqnarray}
|\vec{x})_{M}=\int \frac{d^{3}p}{(2\pi)^{\frac{3}{2}}}e^{-i\vec{p}.\vec{x}}|\vec{p})\nonumber\\
=\frac{\theta^{3/4}}{2\pi}\int \frac{d^{3}p}{(2\pi)^{\frac{3}{2}}}e^{i\vec{p}.(\hat{\vec{x}}-\vec{x})}
\label{mortho1}
\end{eqnarray}
satisfying 
\begin{eqnarray}
\int d^3x |\vec{x})_M\star_M{}_M(\vec{x}|&=&\int d^3x|\vec{x})_M{}_M(\vec{x}|=1_q,\nonumber\\
\label{mcomplt1}
(\vec{p}|\vec{x})_M&=&\frac{1}{(2\pi)^{\frac{3}{2}}}e^{-i\vec{p}\cdot \vec{x}},\nonumber\\
{}_M(\vec{x}|\vec{x}^\prime)_M&=&\delta^3(\vec{x}-\vec{x}^\prime ),
\label{moyalcf}
\end{eqnarray} 
where 
\begin{eqnarray}
\star_{M}=e^{\frac{i}{2}\theta^M_{ij}\overleftarrow\partial_{i}\overrightarrow\partial_{j}}
\label{moystar}
\end{eqnarray}
with $\theta^M_{ij}=\theta_{ij}$.  These basis states are simultaneous eigenstates of $\hat{X_{i}}^{(c)}\forall i:{}_M(\vec{x}|\hat{X_{i}}^{(c)}|\vec{x}^\prime)_M= x_{i}\delta^3(\vec{x}-\vec{x}^\prime )$.

As before we can impose the additional structure of an algebra on $\mathcal H_{q}^{(3)}$ by defining the multiplication map
\begin{eqnarray}
m(|\psi)\otimes |\phi))=|\psi\phi),
\label{map}
\end{eqnarray}
where normal operator multiplication is implied on the right. Expanding a pair of generic states states $|\psi)$ and $|\phi)$ as
\begin{eqnarray}
|\psi)=\frac{\theta^{3/4}}{2\pi}\int \frac{d^{3}p}{(2\pi)^{3/2}}\psi(\vec{p})e^{ip_{i}\hat{x}_{i}}\nonumber\\
=\frac{\theta^{3/4}}{2\pi}\int \frac{d^{3}\bar{p}}{(2\pi)^{3/2}}\psi(\vec{\bar{p}})e^{i\bar{p}_{i}\hat{\bar{x}}_{i}},
\end{eqnarray}
and like-wise for $|\phi)$, a straightforward computation yields
\begin{eqnarray}
_{M}(\vec{x}|\psi\phi)=2\pi\theta^{3/4}\;_{M}(\vec{x}|\psi)\star_{M}{}_{M}(\vec{x}|\phi)
\label{moy4}
\end{eqnarray}
where 
\begin{equation}
_M(\vec x|\psi)=\int \frac{d^{3}p}{(2\pi)^{3/2}}\psi(\vec{p})e^{ip_{i}x_{i}}
\end{equation}
and $\star_{M}=e^{\frac{i}{2}\bar{\theta}_{\alpha \beta}\overleftarrow{\bar{\partial}}_{\alpha}\overrightarrow{\bar{\partial}}_{\beta}}$ has been written in the barred coordinates $\bar{x}_{i}$. However, as $\theta_{ij}\overleftarrow\partial_{i}\overrightarrow\partial_{j}$ is a $SO(3)$ scalar bi-differential operator, we can readily switch to our fiducial c-number coordinates $x_{i}=(\bar{R}_{ij})\bar{x}_{j}$ to yield the Moyal star product (\ref{moystar}). 

Next we introduce the Voros basis in three dimensions through an expansion in momentum basis in complete analogy with the two dimensional case:
\begin{eqnarray}
|x)_{V}= \frac{\theta^{3/4}}{\sqrt{2\pi}}\int d^{3}p e^{-\frac{\theta \vec{p}^{2}}{4}}e^{-i\vec{p}\cdot \vec{x}}|p)=\int \frac{d^{3}p\theta^{3/2}}{(2\pi)^{3/2}}e^{-\frac{\theta \vec{p}^{2}}{4}}e^{i\vec{p}\cdot(\hat{\vec{x}}-\vec{x})}.
\label{3dvoros}
\end{eqnarray}
These states satisfy
\begin{eqnarray}
&&\int \frac{d^3x}{(2\pi)^2\theta^{3/2}}|\vec{x})_V\star_V\;_V(\vec{x}|=1_q,\\
&&_V(x|p)=\frac{\theta^{3/4}}{\sqrt{2\pi}}e^{-\frac{\theta \vec{p}^{2}}{4}}e^{i\vec{p}\cdot \vec{x}},\\
&&_V(\vec{x}^\prime|\vec{x})_V=\sqrt{2\pi}e^{-\frac{1}{2\theta}(\vec{x}-\vec{x}^\prime)^2},
\end{eqnarray}
where
\begin{eqnarray}
\star_{V}=e^{\frac{i}{2}\theta^V_{ij}\overleftarrow\partial_{i}\overrightarrow\partial_{j}}
\label{vorstar}
\end{eqnarray}
with $\theta^V_{ij}=-i\theta\delta_{ij}+\theta_{ij}$ and $\theta$ as in \ref{thetab}. 

A straightforward calculation now yields
\begin{eqnarray}
_{V}(\vec{x}|\psi\phi)=_{V}(\vec{x}|\psi)\star_{V} {}_{V}(\vec{x}|\phi)
\label{vor4}
\end{eqnarray}
where 
\begin{equation}
_V(\vec x|\psi)=\frac{\theta^{3/4}}{\sqrt{2\pi}}\int \frac{d^{3}p}{(2\pi)^{3/2}}\psi(\vec{p})e^{-\frac{\theta \vec{p}^{2}}{4}}e^{ip_{i}x_{i}}.
\end{equation}

\section{Angular Momentum Operator}
\label{angmom}
Next we construct the representation of the angular momentum operator (generators of rotations) on the quantum Hilbert space. Consider the momentum basis expansion of a state $|\psi)$ in the quantum Hilbert space
\begin{eqnarray}
|\psi)=\psi(\hat{\vec{x}})=\int d^{3}p \psi(\vec{p})e^{ip_{i}\hat{x}_{i}}.
\end{eqnarray}

Let us consider an arbitrary infinitesimal rotation $R \in SO(3)$, which rotate the coordinate system: $\hat{x}_{i}\rightarrow \hat{x}_i^R=R_{ij}\hat{x}_{j}$ with
\begin{eqnarray}
R=1+i\vec{\phi}\cdot\vec{L}=1+i\phi_{i}L_{i}.
\label{inf}
\end{eqnarray}
Here the $L_i$'s are the 3 dimensional matrix representations of the $SO(3)$ generators and $\vec{\phi}$ the infinitesimal rotational parameters.  We are looking for the infinitesimal unitary operator that implements this transformation on the quantum Hilbert space.  Keeping in mind the scalar nature of the 'wave functions' one wants 
\begin{eqnarray}
|\psi^{R})=U({R})|\psi)=\psi^{R}(\hat{\vec{x}})=\psi(R^{-1}\hat{\vec{x}})&=&\int d^{3}p \psi(\vec{p})e^{ip_{i}(R^{-1}\hat{\vec{x}})_{i}}.\nonumber
\end{eqnarray}
Using (\ref{inf}), this can be recast in the form
\begin{eqnarray}
\psi^{R}(\hat{\vec{x}})&=&\int d^{3}p\psi(\vec{p})e^{ip_{i}(\delta_{ij}-i\phi_{l}(L_{l})_{ij})\hat{x}_{j}}\nonumber\\
&=&\int d^{3}p \psi(\vec{p}) e^{ip_{i}\hat{x}_{i}+\phi_{l}(L_{l})_{ij}\hat{x}_{j}p_{i}}.\nonumber\\
\end{eqnarray}
Using the Baker-Campbell-Hausdorff formula and retaining terms up to linear order in $\phi_{l}$, we can write this as
\begin{eqnarray}
\psi^{R}(\hat{\vec{x}})&=&\int d^{3}p\psi(\vec{p})[1+\phi_{l}(L_{l})_{ij}\hat{x}_{j}p_{i}+\frac{1}{2}\phi_{l}p_{i}(L_{l})_{ij}p_{m}\theta_{jm}]e^{ip_{i}\hat{x}_{i}}\nonumber\\
&=& (1+\phi_{l}(L_{l})_{ij}\hat{X}^{(c)}_{j}\hat{P}_{i})\psi(\hat{\vec{x}}).
\end{eqnarray}
Using the explicit forms of the $SO(3)$ generators $\vec{L}$
\begin{eqnarray}
(L_{i})_{jk}= -i \epsilon_{ijk};[L_{i},L_{j}]=i\epsilon_{ijk}L_{k},
\label{ang}
\end{eqnarray}
one gets
\begin{eqnarray}
\label{infrot}
 \psi^{R}(\hat{x}_{i})=\psi(\hat{x}_{i})+ i\phi_{i}\hat{J}_{i}\psi(\hat{x})
\end{eqnarray}
from which we identify 
\begin{eqnarray}
\hat{J}_{i}=\epsilon_{ijk}\hat{X}^{(c)}_{j}\hat{P}_{k}
\label{toang}
\end{eqnarray}
as the generators of rotations, i.e., the angular momentum operators acting on $\mathcal H^{(3)}_{q}$.  It can be easily checked that the operators $J_{i}$ satisfy the standard  $SO(3)$ commutation relations
\begin{eqnarray}
[\hat{J}_{i},\hat{J}_{j}]=i\epsilon_{ijk}\hat{J}_{k},
\end{eqnarray}
and furnishes a representation of $L_{i}$ on the quantum Hilbert space.  As usual the operator $U(R)$ for a finite rotation with rotation parameters $\vec\phi$ is given by $U(R)=e^{i\vec\phi\cdot\hat{\vec{J}}}$ and is unitary as can be easily verified. 

The angular momentum, unlike the linear momentum $\hat{P}_{i}$, does not satisfy the usual Leibniz rule. While the Leibniz rule for $\hat{P}_{i}$ is trivial, it is also not difficult to see that it is not satisfied for $\hat{J}_{i}$ by considering its action on an arbitrary product state $(\phi\psi)$:
\begin{eqnarray}
\hat{J}_{i}(\phi\psi)= \epsilon_{ijk}\hat{X}^{(c)}_{j}\hat{P}_{k}(\phi\psi)=\epsilon_{ijk} \hat{X}^{(c)}_{j}((\hat{P}_{k}\phi)\psi +\phi(\hat{P}_{k}\psi)).
\end{eqnarray}
Given the definition of $\hat{X}^{(c)}_{j}$, which is really the average of left and right actions, this simply cannot be written as 
\begin{eqnarray}
(\epsilon_{ijk}\hat{X}^{(c)}_{j}(\hat{P}_{k}\phi))\psi +\phi(\epsilon_{ijk}\hat{X}^{(c)}_{j}(\hat{P}_{k}\psi)).
\end{eqnarray}
Rather, it should be written as
\begin{eqnarray}
\hat{J}_{i}(\phi\psi)=\frac{1}{2}\epsilon_{ijk}[\hat{x}_{j}((\hat{P}_{k}\phi)\psi +\phi(\hat{P}_{k}\psi))+((\hat{P}_{k}\phi)\psi +\phi(\hat{P}_{k}\psi))\hat{x}_{j})].
\end{eqnarray}
This generates factors like $(\hat{x}_{j}\phi)$ and $(\psi\hat{x}_{j})$. Expressing them as $([\hat{x}_{j},\phi]+\phi \hat{x}_{j})$ and $([\psi,\hat{x}_{j}]+\hat{x}_{j}\psi)$, respectively, and substituting them back in the above equation yields, after some re-arrangement,
\begin{eqnarray}
\label{defleib}
\hat{J}_{i}(\phi\psi)=(\hat{J}_{i}\phi)\psi +\phi(\hat{J}_{i}\psi) +\frac{1}{2}\epsilon_{ijk}([\hat{x}_{j},\phi](\hat{P}_{k}\psi)-(\hat{P}_{k}\phi) [\hat{x}_{j},\psi]).
\label{actang}
\end{eqnarray}
Clearly the first two terms here corresponds to what one expects from the naive Leibniz rule and the third term represents the corresponding modification/deformation.

We can now use the identity
\begin{eqnarray}
[\hat{x}_{i},\psi]=-\theta_{ij}(\hat{P}_{j}\psi),
\end{eqnarray}
which follows from (\ref{antisym}-\ref{comm}), to re-express (\ref{actang}) in terms of $\vec{\theta}$ as
\begin{eqnarray}
\label{Leibnitz}
\hat{J}_{i}(\phi\psi)=(\hat{J}_{i}\phi)\psi +\phi(\hat{J}_{i}\psi)+ \frac{1}{2}[(\hat{P}_{i}\phi)((\vec{\theta}\cdot\vec{P})\psi)-((\vec{\theta}\cdot\vec{P})\phi)(\hat{P}_{i}\psi)].
\end{eqnarray}

\section{Necessity of deformed co-product to restore the automorphism symmetry under $SO(3)$ rotations }
\label{autosec}
The generic state $|\psi)$ transforms under a SO(3) rotation ${R}$ as
\begin{eqnarray}
|\psi)\rightarrow |\psi^{R})&=& \int d^{3}p \psi(\vec{p}) e^{i\vec{p}\cdot(R^{-1}\hat{\vec{x}})}\\
&=& \int d^{3}p \psi(\vec{p}) e^{i(R\vec{p})\cdot\hat{\vec{x}}}.\nonumber\\
\label{genstate}
\end{eqnarray}
Likewise we can introduce another state 
\begin{eqnarray}
|\phi)=\int d^{3}p \phi(\vec{p})e^{i\vec{p}\cdot\hat{\vec{x}}}
\end{eqnarray}
and its rotated counterpart
\begin{eqnarray}
|\phi^{R})= \int d^{3}p \phi(\vec{p}) e^{i(R\vec{p})\cdot\hat{\vec{x}}}.
\end{eqnarray}
The aforementioned structure of the algebra (\ref{map}) allows us to write the composite state $|\psi \phi)$ as
\begin{eqnarray}
|\psi\phi)=\int d^{3}p d^{3}p' \psi(\vec{p}) \phi(\vec{p'})e^{i(\vec{p}+\vec{p'})\cdot\hat{\vec{x}}}e^{-\frac{i}{2}p_{i}p'_{j}\theta_{ij}}
\label{state1}
\end{eqnarray}
where we have made use of the Baker-Campbell-Hausdorff formula. 

Consider the rotated state $|(\psi\phi)^{R})$, which is obtained by applying the rotation ${R}\in SO(3)$ on the state $|\psi\phi)$. Clearly,
\begin{eqnarray}
|(\psi\phi)^{R})&=& U({R})|\psi\phi)= U({R})[m(|\psi)\otimes |\phi))]\nonumber\\
&=&\int d^{3}p d^{3}p'\psi(\vec{p})\phi(\vec{p'})e^{i(R(\vec{p}+\vec{p'})).\hat{\vec{x}}}e^{-\frac{i}{2}p_{i}p'_{j}\theta_{ij}}.
\label{hopf}
\end{eqnarray}
At this stage it can be observed that if it were the case of commutative quantum mechanics, we would have automorphism symmetry i.e we can write
\begin{eqnarray}
|(\psi\phi)^{R})=|\psi^{R}\phi^{R})
\end{eqnarray}
where the RHS can be easily seen to be obtained from the undeformed co-product $\Delta_{0}(R)=U(R)\otimes U(R)$ and the RHS can be expressed as
\begin{eqnarray}
|\psi^{R}\phi^{R})= m[\Delta_{0}(R)(|\psi)\otimes |\phi))].
\end{eqnarray}
However, as we show now this situation changes drastically in the non-commutative case; here we are forced to apply a deformed co-product $\Delta_{\theta}(R)$, which goes over to $\Delta_{0}(R)$ only in the limit $\theta \rightarrow 0$, in order to recover the automorphism symmetry. To this end, consider $m[\Delta_{\theta}(R)(|\psi)\otimes |\phi))]$ with 
\begin{eqnarray}
\Delta_{\theta}(R)=F \Delta_{0}(R)F^{-1}
\label{ftwist}
\end{eqnarray}
and
\begin{eqnarray}
 F=e^{\frac{i}{2}\alpha_{ij}\hat{P_{i}}\otimes \hat{P_{j}}}.
 \label{twistb}
\end{eqnarray}
This ansatz is motivated from earlier studies using the Moyal/Voros basis \cite{Banerjee}. A straight forward calculation yields 
\begin{eqnarray}
& & m[\Delta_{\theta}(R)(|\psi)\otimes |\phi))] \nonumber \\
& & =\int d^{3}p d^{3}p'\psi(\vec{p})\phi(\vec{p'})e^{-\frac{i}{2}\alpha_{mn}p_{m}p'_{n}}e^{\frac{i}{2}\alpha_{kl}(R\vec{p})_{k}(R\vec{p'})_{l}}e^{iR(\vec{p}+\vec{p'}).\hat{\vec{x}}}e^{-\frac{i}{2}(R\vec{p})_{i}(R\vec{p'})_{j}\theta_{ij}}. \nonumber \\
\label{atwist}
\end{eqnarray}
The above expression will coincide with that of $|(\psi\phi)^{R})$ given in (\ref{hopf}) iff the $\alpha$ matrix is identified with the NC matrix $\Theta :\alpha= \Theta$. 

We therefore have the final expression of the twist and the co-product given in an abstract i.e. basis independent form as 
\begin{eqnarray}
F=e^{\frac{i}{2}\theta_{ij}\hat{P_{i}}\otimes\hat{P_{j}}} 
\label{indtwist}
\end{eqnarray}
and
\begin{eqnarray}
 \Delta_{\theta}(R)= F\Delta_{0}(R)F^{-1}.
\end{eqnarray}
These are the essential deformed Hopf-algebraic structures \cite{Aschieri} required to restore the automorphism symmetry:
\begin{equation}
\label{auto}
|(\psi\phi)^R)=U(R)[m(|\psi)\otimes|\phi))]=m[\Delta_\theta(R)(|\psi)\otimes|\phi))].
\end{equation}

The same conclusion can be reached from the deformed Leibnitz rule (\ref{Leibnitz}).  Indeed, the deformed co-product can simply be read off as
\begin{eqnarray}
\bigtriangleup_{\theta}(\hat{J}_{i})=\bigtriangleup_{0}(\hat{J}_{i}) +\frac{1}{2}[\hat{P}_{i}\otimes (\vec{\theta}\cdot\vec{P})- (\vec{\theta}\cdot\vec{P})\otimes \hat{P}_{i}],
\label{deformed}
\end{eqnarray}
with $\triangle_{0}(\hat{J}_{i})=\hat{J}_{i}\otimes 1 +1\otimes \hat{J}_{i}$ the undeformed co-product for $SO(3)$ generators.  It can easily be checked that this is precisely the deformed co-product obtained from
\begin{eqnarray}
\bigtriangleup_{\theta}(\hat{J}_{i})= F \bigtriangleup_{0}(\hat{J}_{i})F^{-1}
\end{eqnarray}
by using the Hadamard identity for the form of the twist (\ref{indtwist}).

Having obtained the abstract form of the twist (\ref{indtwist}), the form of the twist $F_{M/V}$ in Moyal/Voros basis can be read off from the overlaps $_{M/V}(\vec{x}|\psi\phi)$ in (\ref{moy4}) and (\ref{vor4}) satisfying
\begin{eqnarray}
_{M/V}(\vec{x}|\psi\phi)=m_{0}[F^{-1}_{M/V}(_{M/V}(\vec{x}|\psi)\otimes _{M/V}(\vec{x}|\phi))]
\end{eqnarray}
where $m_{0}$ represents the point-wise multiplication map:
\begin{eqnarray*}
m_{0}[\psi(\vec{x})\otimes \phi(\vec{x})]= \psi(\vec{x})\phi(\vec{x}).
\end{eqnarray*}
This yields,
\begin{eqnarray}
&&F_{M}=e^{-\frac{i}{2}\theta^M_{ij}\partial_{i}\otimes \partial_{j}}=e^{-\frac{i}{2}\theta_{ij}\partial_{i}\otimes \partial_{j}}\\
&&F_{V}=e^{-\frac{i}{2}\theta^{V}_{ij}\partial_{i}\otimes \partial_{j}}=e^{-\frac{i}{2}(-i\theta \delta_{ij}+\theta_{ij})\partial_{i}\otimes \partial_{j}}.
\end{eqnarray}

Finally we remark that the restoration of the SO(3) automorphism symmetry is only relevant in the one particle setting insofar as the action of the one particle potential $V(\hat{X}_i)$ on the quantum state $\psi(\hat{x}_i)$ corresponds to operator multiplication, i.e., $V(\hat{X}_i)\psi(\hat{x}_i)=V(\hat{x}_i)\psi(\hat{x}_i)$ as the action of the quantum position operators $\hat{X}_i$ was defined through left multiplication.  In this setting the composite state $|\psi\phi)$ represents a single particle state that results from the action of an observable that depends on the coordinates $\hat{X}_i$ alone, i.e., $\psi(\hat{X}_i)|\phi)=|\psi(\hat{x}_i)\phi)$ (see \ref{position}).  

In the multi-particle setting the restoration of the automorphism symmetry becomes relevant in the context of the manifest restoration of the symmetry on the level of the action.  If we were to view the Schroedinger equation as a field equation resulting from the action
\begin{equation}
\label{action}
S=\int dt tr_c\psi^\dagger\left(i\partial_t-\frac{\hat{P}^2}{2m}-V(\hat{x}_i)\right)\psi,
\end{equation}
the rotation symmetry is manifest if the Lagrangian density transforms as a scalar under rotations, i.e., $(\psi^\dagger\psi)\rightarrow (\psi^\dagger\psi)^R$ and $(\psi^\dagger V\psi)\rightarrow (\psi^\dagger V\psi)^R$ since $tr_c(A)^R=tr_c A$ for a generic composite $A$ of fields (see section\ref{consangmom}). To achieve this it is necessary to implement the deformed co-product on the composites of fields appearing in the Lagrangian, as explained in \cite{Pinzul}. We return to the transformation properties of the Schroedinger action (\ref{action}) in section \ref{consangmom}. 

The same considerations apply to relativistic non-commutative field theories where the restoration of the Lorentz automorphism symmetry is required to make the Lagrangian manifestly Lorentz invariant.

\section{On the constancy of $\Theta$}
\label{theta}
It is a simple matter to verify that the commutation relations satisfied by the rotated coordinate operator $\hat{x}^{R}_{i}\equiv (R\hat{\vec{x}})_{i}$ are given by 
\begin{equation}
[\hat{x}^{R}_{i},\hat{x}^{R}_{j}]= \hat{x}^{R}_{i}\hat{x}^{R}_{j}-\hat{x}^{R}_{j}\hat{x}^{R}_{i}=i(R\Theta R^{T})_{ij}\equiv i\left(\Theta_{UD}\right)_{ij}
\end{equation}
Clearly, here the $\Theta$ matrix transforms as a second rank antisymmetric tensor under rotations $R\in SO(3)$.  This is actually due to fact that we had implicitly used the undeformed co-product, reflected by the notation $\Theta_{UD}$, to compute the commutator:
\begin{eqnarray}
[\hat{x}^{R}_{i},\hat{x}^{R}_{j}]= m[\Delta_{0}(R)(\hat{x}_{i}\otimes \hat{x}_{j}-\hat{x}_{j}\otimes \hat{x}_{i})],
\end{eqnarray}
which is nothing but the commutator of the rotated coordinate operators.  However, as we have seen in the previous section, we should really use the deformed co-product $\Delta_{\theta}$ to compute the rotated commutator, as the composite object $(\hat{x_{i}}\hat{x_{i}})$  transforms to $(\hat{x_{i}}\hat{x_{j}})^{R}$ under rotation by $R\in SO(3)$, and this will be implemented by the deformed co-product $\triangle_{\theta}(R)$. In other words we have to simply replace $\Delta_{0}(R)$ by $\Delta_{\theta}(R)$ in the above computation. With this the above commutator gets replaced by
\begin{eqnarray}
([\hat{x}_{i},\hat{x}_{j}])^{R}=(\hat{x}_{i}\hat{x}_{j})^{R}-(\hat{x}_{j}\hat{x}_{i})^{R}= m[\Delta_{\theta}(R)(\hat{x}_{i}\otimes \hat{x}_{j}-\hat{x}_{j}\otimes \hat{x}_{i})].
\end{eqnarray}
Now a straightforward computation yields
\begin{eqnarray}
(\hat{x}_{i}\hat{x}_{j})^{R}&=& m[\Delta_{\theta}(R)(\hat{x}_{i}\otimes \hat{x}_{j})]\nonumber\\
&=& \hat{x}^{R}_{i}\hat{x}^{R}_{j}+\frac{i}{2}\theta_{ij}-\frac{i}{2}(\Theta^{R}_{UD})_{ij}.
\label{nonco}
\end{eqnarray}
This indicates that the composite object no longer transforms as a second rank tensor under rotation. However a simple antisymmetrization now yields
\begin{eqnarray}
(\hat{x}_{i}\hat{x}_{j})^{R}-(\hat{x}_{j}\hat{x}_{i})^{R}= i(\Theta_{D})_{ij}=i\theta_{ij}.
\label{nonco1}
\end{eqnarray}
This implies that the rotation applied to the commutator as a whole is different from the commutator of rotated coordinates.
Here $\Theta_{D}$ refers to the fact that the deformed co-product has been used for its computation. It shows that the NC matrix $\Theta$ really remains invariant under spatial rotations and is the same as in the fiducial frame if the proper deformed co-product action is considered. This deformed co-product on the other hand  arises from the demand of restoration of automorphism symmetry under rotation, as we have seen earlier.

The above result suggests that the commutator $[\hat{x}_j,\hat{x}_k]$ should be invariant under SO(3) transformations.  This is a useful consistency check as one can indeed verify explicitly that
\begin{eqnarray}
\hat{J}_{i}\theta_{jk}&=&-i \hat{J}_{i}[m(\hat{x}_{j}\otimes \hat{x}_{k}-\hat{x}_{k}\otimes\hat{x}_{j})]\nonumber\\
&=& -im[\bigtriangleup_{\theta}(\hat{J}_{i})(\hat{x}_{j}\otimes \hat{x}_{k}-\hat{x}_{k}\otimes \hat{x}_{j})]\nonumber\\
&=& 0,
\label{deform}
\end{eqnarray}
again displaying the constancy of $\theta_{ij}$ under the action of the deformed co-product.

The same conclusion can be drawn from a more general setting, where transformation properties under rotation of the quantum position operator, (introduced earlier in (\ref{op3}),(\ref{op4})) in conjuction with an arbitrary state $\psi(\hat{x}_{i})$ is considered.

To that end, recall that the quantum position operator act from the left on the elements of the quantum Hilbert space
\begin{eqnarray}
\hat{X}^{(l)}_{i}:\psi(\hat{x}_{i})\rightarrow \hat{x}_{i} \psi(\hat{x}_{i})=m(\hat{x}_{i}\otimes \psi(\hat{x}_{i})),
\end{eqnarray}
i.e,
\begin{eqnarray}
\hat{X}^{(l)}_{i}|\psi(\hat{x}_{i}))= |\hat{x}_{i}\psi(\hat{x}_{i})).
\end{eqnarray}
More generally, one may introduce the rotated quantum position operator as the map
\begin{eqnarray}
\hat{X}^{(l){R}}_{i}:\psi(\hat{x}_{i})\rightarrow \hat{x}^{R}_{i}\psi(\hat{x}_{i})=m(\hat{x}^{R}_{i}\otimes \psi(\hat{x}_{i})),
\end{eqnarray}
i.e,
\begin{eqnarray}
\hat{X}^{(l){R}}_{i}|\psi(\hat{x}_{i}))=|\hat{x}^{R}_{i}\psi(\hat{x}_{i}))=R_{ij}\hat{X}^{(l)}_{j}|\psi(\hat{x}_{i})).
\end{eqnarray}
These are straightforward extensions of $\hat{x}_{i}$ and $\hat{x}^{R}_{i}$, acting on $\mathcal H^{(3)}_{c}$, to position observables, acting on $\mathcal H^{(3)}_{q}$, and on their own transform covariantly under rotations. One can, however, ask what is the transformation property of the composite object $(\hat{X}^{(l)}_{i}|\psi(\hat{x}_{i}))\rightarrow (\hat{X}^{(l)}_{i}|\psi(\hat{x}_{i}))^{R}$ under a rotation. Unlike the commutative case, we can expect a deformation of a vectorial nature through our experience of the non-covariant transformation property $(\hat{x}_{i}\hat{x}_{j})\rightarrow (\hat{x}_{i}\hat{x}_{j})^{R}$ in (\ref{nonco}). To show that this is indeed the case, observe that
under a rotation, $m(\hat{x}_{i}\otimes \psi(\hat{x}_{i}))\rightarrow U(R)[m(\hat{x}_{i}\otimes \psi(\hat{x}_{i}))]= m[\Delta_{\theta}(R)(\hat{x}_{i}\otimes \psi(\hat{x}_{i}))]$.  A straightforward computation now yields, on using (\ref{nonco1}) (see Appendix),
\begin{eqnarray}
m[\Delta_{\theta}(R)(\hat{x}_{i}\otimes \psi(\hat{x}_{i}))]=\hat{\tilde{X}}^{(l){R}}_{i}|\psi^{R}(\hat{x}_{i})),
\label{btwist}
\end{eqnarray}
where
\begin{eqnarray}
\hat{\tilde{X}}^{(l){R}}_{i}\equiv \hat{X}^{(l)^{R}}_{i}+\frac{1}{2}[R,\Theta]_{ij}\hat{P}_{j}
\label{rotated}
\end{eqnarray}
 can be regarded as the {\it effective} rotated quantum position operator and is distinguished by a tilde. It can now be trivially checked that
\begin{eqnarray}
[\hat{\tilde{X}}^{(l){R}}_{i},\hat{\tilde{X}}^{(l){R}}_{j}]=i\theta_{ij} \neq [\hat{x}^{R}_{i},\hat{x}^{R}_{j}]
\label{nonalg}
\end{eqnarray}
again showing the constancy of $\Theta$. The eqs.(\ref{rotated},\ref{nonalg}) reproduce the result of \cite{Castro}, obtained in a Hopf-algebraic framework and furnishes a derivation from a somewhat different perspective of the result obtained in \cite{Wess},\cite{Chaichian}.
At this stage we would like to make some pertinent observations:\\
(i) Note that the expression (\ref{rotated}) has been obtained in a self-consistent approach, as the invariance of $\Theta$ in the sense of (\ref{nonco1}) has been made use of here. For the special case $\psi(\hat{x}_{i})=\hat{x}_{j}^{R}$, one can easily show, by using (\ref{nonco}), that
\begin{eqnarray}
\hat{\tilde{X}}_{i}^{(l){R}}\hat{x}_{j}^{R}=(\hat{x}_{i}\hat{x}_{j})^{R}.
\end{eqnarray}
(ii) The distinction between $\hat{X}^{(l){R}}_{i}$ and $\hat{\tilde{X}}^{(l){R}}_{i}$ disappear in two spatial dimension as $[R,\Theta]=0$ identically.\\
(iii) It also disappears in the case of no rotation i.e $R=1$\\
(iv) The form of $\hat{\tilde{X}}^{(l){R}}_{i}$ (98) shows that $\hat{X}^{(l)}_{i}$ does not transform covariantly, for $D> 2$ if the transformation property $\hat{X}^{(l)}_{i}$ in conjunction with a state is considered. This is the price one has to pay to hold $\Theta$ fixed. In other words, this non-covariant transformation property is induced on it by the deformed co-product, when the composite object $(\hat{x}_{i}\psi(\hat{x}_{i}))$ undergoes rotation. In this sense, $\hat{\tilde{X}}^{(l){R}}_{i}$ does not enjoy a fundamental status like $\hat{X}^{(l){R}}_{i}$.\\
(v) Proceeding similarly, the transformation property of $\hat{X}^{(r)}_{i}$, corresponding to right action is found to be
\begin{eqnarray}
\hat{\tilde{X}}^{(r){R}}_{i}=\hat{X}^{(r){R}}_{i}-\frac{1}{2}[R,\Theta]_{ij}\hat{P}_{j},
\end{eqnarray}
so that $\hat{X}^{(c)}_{i}$ in (\ref{aver}) transform covariantly
\begin{eqnarray}
\hat{X}^{(c)}_{i}\rightarrow \hat{X}^{(c){R}}_{i}=R_{ij}\hat{X}^{(c)}_{j}.
\end{eqnarray}

\section{$SO(3)$ transformation properties of the Shroedinger action and Hamiltonian}
\label{consangmom}

Let us consider the motion of a particle described by the Hamiltonian
\begin{eqnarray}
\label{ham}
H=\frac{\vec{P}^{2}}{2m}+V(\hat{X}_{i}),
\label{har}
\end{eqnarray}
where $V(\hat{X}_{i})$ represents the potential. Note that the argument of $V(\hat{X}_{i})$ is $\hat{X}_{i}\equiv \hat{X}^{(l)}_{i}$ and not $\hat{x}_{i}$ since, like the kinetic energy term, it is a operator on $\mathcal H^{(3)}_{q}$. 

In particular we focus on rotational invariant potentials in the conventional sense: 
\begin{eqnarray}
\label{rotinv}
V(\hat{X}^{R}_{i})=V(\hat{X}_{i}),
\label{pot1}
\end{eqnarray}
where $\hat{X}^{R}_{i}=R_{ij}\hat{X}_{j}$.  An example of such a rotational invariant potential that we study later is the isotropic harmonic potential
\begin{eqnarray}
V(\hat{X}_{i})=\frac{1}{2}m\omega^{2}\hat{X}_{i}\hat{X}_{i}.
\label{pot}
\end{eqnarray}

As the action of the position operators is defined through left multiplication, the action of the potential on the quantum state is 
\begin{eqnarray}
V(\hat{X}_{i})\psi(\hat{x}_{i})=V(\hat{x}_{i})\psi(\hat{x}_{i}).
\label{pot2}
\end{eqnarray}
In the above the potential $V(\hat{x}_i)$ is also an operator acting on the classical configuration space.  This operator acts on the quantum state through ordinary operator multiplication.  In the light of (\ref{rotinv}) this operator satisfies (see also (\ref{infrot}))
\begin{equation}
\label{rotinv1}
\hat{J}_{i}V(\hat{x}_{i})=0,
\end{equation}
which can indeed be explicitly verified for the harmonic oscillator potential.

Let us now consider the issue of rotational invariance of the Hamiltonian (\ref{ham}).  It is a simple matter to see that the angular momentum operators $J_i$ commute with the kinetic energy term, but not with the potential $V(\hat{X}_i)$, even though the potential is rotational invariant in the sense of (\ref{rotinv}).  It is useful to understand the origin of this non-commutativity more precisely in the context of the deformed Leibnitz rule (\ref{Leibnitz}).  For this purpose let us consider 
\begin{equation}
\label{L1}
J_iV(\hat{X}_{i})\psi(\hat{x}_{i})=[J_i,V(\hat{X}_{i})]\psi(\hat{x}_{i})+V(\hat{X}_{i})J_i\psi(\hat{x}_{i}).
\end{equation}
On the other hand from (\ref{Leibnitz}) this can also be written as
\begin{eqnarray}
\label{L2}
J_iV(\hat{X}_{i})\psi(\hat{x}_{i})=&&J_i\left(V(\hat{x}_{i})\psi(\hat{x}_{i})\right)=\left(J_iV(\hat{x}_{i})\right)\psi(\hat{x}_{i})+V(\hat{x}_{i})\left(J_i\psi(\hat{x}_{i})\right)\nonumber\\
&&+\frac{1}{2}[(\hat{P}_{i}V)((\vec{\theta}\cdot\vec{P})\psi)-((\vec{\theta}\cdot\vec{P})V)(\hat{P}_{i}\psi)].
\end{eqnarray}
For rotational invariant potentials the first term on the right of (\ref{L2}) vanishes by (\ref{rotinv1}) and since $\psi$ is an arbitrary state we conclude
\begin{equation}
\label{comm1}
[J_i,V(\hat{X}_{i})]=\frac{1}{2}[(\hat{P}_{i}V)\vec{\theta}\cdot\vec{P}-((\vec{\theta}\cdot\vec{P})V)\hat{P}_{i}].
\end{equation}

We note that even for rotational invariant potentials the Hamiltonian and angular momentum do not commute and that the rotational symmetry is explicitly broken on the level of the Hamiltonian.  However, the non-vanishing commutator originates purely from the deformation of the Leibnitz rule, which vanishes in the commutative limit.  Hence the breaking of rotational symmetry on the level of the Hamiltonian for rotational invariant potentials results purely from the deformation.  Another way of phrasing this statement is to note that $(V\psi)^R\ne V^R\psi^R$, while the equality is required to make the rotational invariance manifest on the level of the Hamiltonian for rotational invariant potentials.  Indeed, $(V\psi)^R=U(R)VU(R)^{-1}\psi^R\equiv V_{eff}^R\psi^R$ where $V_{eff}^R$ can be thought of as an effective potential in the rotated frame.  For infinitesimal rotations the form of $V_{eff}^R$ can be read of from (\ref{comm1}).

Let us repeat the above analysis on the level of the Schroedinger action (\ref{action}).  We note that this action is invariant under the following transformation:
\begin{eqnarray}
\label{trans1}
\psi^\dagger&&\rightarrow (\psi^\dagger)^R=U(R)\psi^\dagger,\nonumber\\  
\psi&&\rightarrow \psi^R=U(R)\psi,\nonumber\\
V\psi&&\rightarrow (V\psi)^R=m[\Delta_\theta(R)(V\otimes \psi)].
\end{eqnarray}   
Note that the first and second equation are not inconsistent as the hermitian conjugation here ($\dagger$) refers to hermitian conjugation on the classical configuration space and not on quantum Hilbert space.  This is analogues to commutative quantum mechanics where complex conjugation (hermitian conjugation here) commutes with rotations.  

To verify the invariance of the action under (\ref{trans1}) as well as the statements above explicitly,  note that $tr_c(\psi^\dagger\phi)=(\psi,\phi)$. Here and in what follows $\phi$ denotes any composite of fields, particularly $V\psi$ in the case of (\ref{trans1}).  Writing 
\begin{equation}
\hat{J}_i\psi^\dagger=\frac{1}{2}\epsilon_{ijk}(\hat{x}_j(\hat{P}_k\psi^\dagger)+(\hat P_k\psi^\dagger)\hat{x}_j),
\end{equation}
one easily verifies from (\ref{act1})
\begin{equation}
(\hat{J}_i\psi^\dagger)^\dagger=-\hat{J}_i\psi,
\end{equation}
which implies
\begin{equation}
(U(R)\psi^\dagger)^\dagger=(e^{i\vec\phi\cdot\hat{\vec{J}}}\psi^\dagger)^\dagger=U(R)\psi.
\end{equation}
Thus one finds
\begin{equation}
\label{trace}
tr_c((\psi^\dagger)^R\phi^R)=(U(R)\psi,U(R)\phi)=(\psi,\phi)=tr_c(\psi^\dagger\phi),
\end{equation}
where we have used the unitarity of $U(R)$ w.r.t. the inner product on the quantum Hilbert space, i.e., $U(R)^\ddagger=U(R)^{-1}$. 

Note that we have actually used the undeformed co-product in the above argument, i.e., we did not apply the deformed co-product to the composite $\psi^\dagger\phi$ in (\ref{trace}).  The same result can, however, be obtained from the deformed co-product as the deformation is essentially irrelevant when considering any term in the action as a product of two composites.  The deformation only manifests itself on the level of the transformation properties of the individual composites.  This follows by first noting from (\ref{act1}) that
\begin{equation}
\label{trace1}
tr_c(\hat{J}_i(\psi^\dagger\phi))=\epsilon_{ijk}\bar{R}_{kl}^{-1}\Gamma_{l\alpha}tr_c([\hat{\bar{x}}_\alpha,\hat{X^{(c)}}_j(\psi^\dagger\phi)])=0,
\end{equation}
since the trace of a commutator, under the trace class condition that ensures that the trace is well defined, vanishes. Here $\psi$ and $\phi$ denote any two composites.  This immediately implies for finite rotations $tr_c(U(R)(\psi^\dagger\phi))=tr_c((\psi^\dagger\phi)^R)=tr_c(\psi^\dagger\phi)$.  From (\ref{defleib}) the LHS of (\ref{trace1}) can also be expressed as
\begin{equation}
\label{trace2}
tr_c(\hat{J}_i(\psi^\dagger\phi))=tr_c((\hat{J}_{i}\psi^\dagger)\phi +\psi^\dagger(\hat{J}_{i}\phi) +\frac{1}{2}\epsilon_{ijk}([\hat{x}_{j},\psi^\dagger](\hat{P}_{k}\phi)-(\hat{P}_{k}\psi^\dagger) [\hat{x}_{j},\phi])).
\end{equation}
The deformation (last term) on the RHS can again be expressed as a total commutator by using (\ref{defleib}) and hence its trace vanishes:
\begin{eqnarray}
&&\epsilon_{ijk}([\hat{x}_{j},\psi^\dagger](\hat{P}_{k}\phi)-(\hat{P}_{k}\psi^\dagger) [\hat{x}_{j},\phi])=\nonumber\\
&&\frac{1}{\theta}\epsilon_{ijk}\bar{R}_{k\ell}^{-1}\Gamma_{\ell\alpha}([\hat{x}_{j},\psi^\dagger][\hat{\bar{x}}_{\alpha},\phi]-[\hat{\bar{x}}_{\alpha},\psi^\dagger] [\hat{x}_{j},\phi])=\nonumber\\
&&\frac{1}{\theta}\epsilon_{ijk}\bar{R}_{k\ell}^{-1}\Gamma_{\ell\alpha}([\hat{x}_{j},\psi^\dagger[\hat{\bar{x}}_{\alpha},\phi]-[\hat{\bar{x}}_{\alpha},\psi^\dagger[\hat{x}_{j},\phi]).
\end{eqnarray}
In the last step the Jacobi identity and the fact that the commutators $[\hat{x}_j,\hat{\bar{x}}_\alpha]$ are constants were used. This shows that the deformation is essentially irrelevant and that (\ref{trace2}) can be written as
\begin{equation}
\label{trace3}
tr_c(\hat{J}_i(\psi^\dagger\phi))=tr_c((\hat{J}_{i}\psi^\dagger)\phi +\psi^\dagger(\hat{J}_{i}\phi)),
\end{equation}
i.e., the Leibnitz rule applies under the trace for the product of two composites.  Note that this is not the case when the product of more than two composites is considered.  For finite rotations this implies that
\begin{equation}
tr_c(\psi^\dagger\phi)=tr_c(\psi^\dagger\phi)^R=tr_c(m(\Delta_\theta(R)(|\psi^\dagger)\otimes|\phi))=tr_c(m(\Delta_0(R)(|\psi^\dagger)\otimes|\phi))=tr_c((\psi^\dagger)^R\phi^R)
\end{equation}
in agreement with (\ref{trace}), which was derived from the undeformed co-product. 

In particular it is worthwhile noting that without any potential, or for any action quadratic in the fields, i.e., $\psi$ and $\phi$ are not composites, the deformation is irrelevant.  This signals that without interactions, i.e., in a single particle description, deformation is not required.  

In the case of (\ref{action}) the deformation only manifests itself through the transformation properties of the composite $\phi=V\psi$ (note that the potential acts as a fixed background field here):
\begin{equation}
(V\psi)^R\ne V^R\psi^R=V\psi^R
\end{equation}
even when $V^R=V$.  This brings us to the same conclusion as in the discussion of the transformation properties of the Hamiltonian, namely, the action is invariant provided that the potential is transformed as follows: $V\rightarrow V_{eff}=U(R)VU(R)^{-1}$ when a rotation is performed, i.e., the action does not preserve its form under rotations.  This applies even to rotational invariant potentials, which are modified under this transformation due to the deformation.  This is in contrast to the commutative case where the action will be form invariant if the potential is rotational invariant.

It is useful to make the considerations above explicit in a soluble example.  This can indeed be done for the $3D$ isotropic harmonic oscillator, described by the potential (\ref{pot}).

As we observed previously, this potential has the $SO(3)$ symmetry, in the sense that it satisfies (\ref{pot1}) and that $\hat{J}_{i}V(\hat{x}_{i})=0$. We can therefore move to the barred frame to write the Hamiltonian as,
\begin{eqnarray}
H=\frac{1}{2m}\vec{\bar{P}}^{2}+\frac{1}{2}m\omega^{2}(\hat{\bar{X}}^{2}_{1}+\hat{\bar{X}}^{2}_{2}+\hat{\bar{X}}^{2}_{3}).
\end{eqnarray}
Recall that in this frame $[\hat{\bar{X}}_{3},\hat{\bar{X}}_{\alpha}]=0$ for $\alpha=1,2$ and $[\hat{\bar{X}}_{1},\hat{\bar{X}}_{2}]=i\theta$. The Hamiltonian can therefore be split into two terms as $H=H_{plane}+H_{line}$ where
\begin{eqnarray}
H_{plane}=\frac{1}{2m}(\hat{\bar{P}}^{2}_{1}+\hat{\bar{P}}^{2}_{2})+\frac{1}{2}m\omega^{2}(\hat{\bar{X}}^{2}_{1}+\hat{\bar{X}}^{2}_{2})
\end{eqnarray}
and
\begin{eqnarray}
H_{line}=\frac{1}{2m}\hat{\bar{P}}^{2}_{3}+\frac{1}{2}m\omega^{2}\hat{\bar{X}}^{2}_{3}
\end{eqnarray}
represent the Hamiltonians for (non-commutative) planar and $1D$ harmonic oscillators, respectively. While, the ground state for $H_{line}$ is well known, the one for $H_{plane}$ has also been worked out in \cite{Gouba}. The complete ground state, therefore, is simply given by the product of these two and is given by 
\begin{eqnarray}
\Psi_{0}(\hat{\bar{x}}_{i})=e^{\frac{\alpha}{2\theta}(\hat{\bar{x}}^{2}_{1}+\hat{\bar{x}}^{2}_{2})}e^{-\frac{1}{2}m\omega^{2}\hat{\bar{x}}^{2}_{3}},
\end{eqnarray}
where '$\alpha$' occuring in the first factor is defined as in \cite{Gouba}. In the following analysis, we shall not need the explicits forms of these coefficients. What is clear is that $\Psi_{0}(\hat{\bar{x}}_{i})$ has only the $SO(2)$ symmetry around the $\bar{x}_{3}$ axis. Since the state written in the unbarred fudicial frame $\Psi_{0}(\hat{x}_{i})$ is related to this frame by a suitable unitary transformation as the components of angular momenta $\hat{\bar{J}}_{i}$ and $\hat{J}_{i}$ in their respective frames, it is advantageous to carry out the computations in this barred frame.

To begin with, it will be advantageous to split $\Psi_{0}$ into the following form
\begin{eqnarray}
\Psi_{0}=\phi \psi,
\end{eqnarray}
where
\begin{eqnarray}
\phi =e^{\frac{\alpha}{2\theta}\hat{x}_{i}\hat{x}_{i}},\nonumber\\
\psi= e^{\frac{1}{2}\bar{\lambda} \hat{\bar{x}}^{2}_{3}},
\end{eqnarray}
and $\bar{\lambda} =-\frac{\alpha}{\theta}-m\omega^{2}$. The advantage of writing in this form ensures
\begin{eqnarray}
\hat{\bar{P}}_{\alpha}\psi =0 ; \hat{\bar{J}}_{i}\phi =0; \hat{\bar{J}}_{3}\psi =0.
\end{eqnarray}
Remembering that we have to use the deformed co-product appropriate for this barred frame, one can write
\begin{eqnarray}
\hat{\bar{J}}_{i}\Psi_{0}=\hat{\bar{J}}_{i}(\phi \psi)=m [\Delta_{\bar{\Theta}}(\hat{\bar{J}}_{i})(\phi \otimes \psi)].
\end{eqnarray}
As it turns out the co-product of $\hat{\bar{J}}_{3}$ undergoes no deformation:
\begin{eqnarray}
\Delta_{\bar{\Theta}}(\hat{\bar{J}}_{3})= \Delta_{0}(\hat{\bar{J}}_{3})=\hat{\bar{J}}_{3}\otimes 1+ 1\otimes \hat{\bar{J}}_{3}.
\end{eqnarray}
Using this,
\begin{eqnarray}
\hat{\bar{J}}_{3}\Psi_{0}=0
\end{eqnarray}
However, for the other components one finds non-vanishing contributions:
\begin{eqnarray}
\hat{\bar{J}}_{\alpha}\Psi_{0}=\phi(\hat{\bar{J}}_{\alpha}\psi)+\frac{\theta}{2}(\hat{\bar{P}}_{\alpha}\phi)(\hat{\bar{P}}_{3}\psi),
\end{eqnarray}
where we have made use of (\ref{Leibnitz}). Now a straight forward computation yields
\begin{eqnarray}
\hat{\bar{J}}_{\alpha}\Psi_{0}=-i\bar{\lambda}(\epsilon_{\alpha \beta}\bar{x}_{3}\phi \hat{\bar{x}}_{\beta}\psi) +\frac{-i\bar{\lambda}}{2}((1-\cosh \theta\alpha)\epsilon_{\alpha \beta}\hat{\bar{x}}_{\beta}+(i \sinh \theta\alpha) \hat{\bar{x}}_{\alpha})\bar{x}_{3}]\Psi_{0}.
\end{eqnarray}
Thus, unlike the commutative case, the ground state does not correspond to vanishing angular momentum. We can attribute this puzzling feature to the presence of the constant "background" $\Theta$ field, which breaks the isotropicity of $3D$ space i.e the $SO(3)$ symmetry breaking to $SO(2)$ (the residual symmetry $\vec{\theta}$ axis i.e $\bar{x}_{3}$-axis).  Despite the fact that the automorphism symmetry can be restored through the deformed co-product this symmetry is not manifest on the level of the action or Hamiltonian, even for rotational invariant potentials.  These features are explicit in the form of the effective potential that can be computed explicitly for finite rotations:
\begin{eqnarray}
{V}_{eff}^{R}(\hat{X}_{i})=V(\hat{X}_{i})-\frac{1}{2}m\omega^2\theta_{i}(R_{ij}-\delta_{ij})\hat{J}_{j}+\frac{1}{8}m\omega^2[(\vec{\theta}.\hat{\vec{P}})^{2}-(\vec{\theta}.\hat{\vec{P}}^{R})^{2}]
\label{potential}
\end{eqnarray}
  
\section{Conclusions}
\label{conc}

We have discussed the generalization of non commutative quantum mechanics to three spatial dimensions.  Particular attention was paid to the identification of the quantum Hilbert space and the representation of the rotation group on it.  Not unexpectedly it was found that this representation undergoes deformation and that the angular momentum operators no longer obey the Leibnitz rule.  This deformation implies that the action for the Schroedinger equation, in which the potential appears as a fixed background field, and Hamiltonian are no longer invariant under rotations, even for rotational invariant potentials. This is in sharp contrast with the commutative case where rotational symmetry is manifest for rotational invariant potentials.  

\section{Acknowledgments} Support under the Indo-South African research agreement between the Department of Science and Technology, Government of India and the National Research Foundation of South Africa is acknowledged, as well as a grant from the National Research Foundation of South Africa. One of author D.S. thanks the Council of Scientific and Industrial Research(C.S.I.R), Government of India, for financial support.

\section{Appendix}
Here we complete 
\begin{eqnarray}
R[m(\hat{x}_{k}\otimes \psi(\hat{x}_{i}))]=m[\bigtriangleup_{\theta}(R)(\hat{x}_{k}\otimes \psi(\hat{x}_{i}))] 
\label{app}
\end{eqnarray}
Using (\ref{ftwist}),(\ref{twistb}), this can be written as,
\begin{eqnarray}
R[m(\hat{x}_{k}\otimes \psi(\hat{x}_{i}))]=m[F(R\otimes R)F^{-1}(\hat{x}_{k}\otimes \psi(\hat{x}_{i}))]
\end{eqnarray}
Now, the action of $F^{-1}$ on $(\hat{x}_{k}\otimes \psi(\hat{x}_{i}))$ is
\begin{eqnarray}
F^{-1}(\hat{x}_{k}\otimes \psi(\hat{x}_{i}))&=& e^{-\frac{i}{2}\theta_{mn}\hat{P}_{m}\otimes \hat{P}_{n}}(\hat{x}_{k}\otimes \psi(\hat{x}_{i}))\nonumber\\
&=& \hat{x}_{k}\otimes \psi(\hat{x}_{i})-\frac{i}{2}\theta_{mn}(\hat{P}_{m}\hat{x}_{k})\otimes(\hat{P}_{n}\psi(\hat{x}_{i}))
\end{eqnarray}
Since $\hat{P}_{m}$ acts adjointly, the factor $(\hat{P}_{m}\hat{x}_{k})$ occurring in the second term can be replaced by $(-i\delta_{mk}1)$, so that $(R\otimes R)F^{-1}(\hat{x}_{k}\otimes \psi(\hat{x}_{i}))$ can be written as
\begin{eqnarray}
(R\otimes R)F^{-1}(\hat{x}_{k}\otimes \psi(\hat{x}_{i}))= (\hat{x}^{R})_{k}\otimes \psi^{R}(\hat{x}_{i})-\frac{1}{2}\theta_{kn} 1\otimes (\hat{P}_{n}\psi(\hat{x}_{i}))^{R}
\end{eqnarray}
Note that here $R$ does not touch $\theta_{ij}$, as follows from (\ref{nonco1}) and it is fixed by fiducial frame we have chosen.
Finally, acting by $F$ on both sides yields,
\begin{eqnarray*}
F(R\otimes R)F^{-1}(\hat{x}_{k}\otimes \psi(\hat{x}_{i}))=e^{\frac{i}{2}\theta_{ij}\hat{P}_{i}\otimes \hat{P}_{j}}((\hat{x}^{R})_{k}\otimes \psi^{R}(\hat{x}_{i})-\frac{1}{2}\theta_{kn} 1\otimes (\hat{P}_{n}\psi(\hat{x}_{i}))^{R})\nonumber\\
=(\hat{x}^{R})_{k}\otimes \psi^{R}(\hat{x}_{i})-\frac{1}{2}\theta_{kn} 1\otimes (\hat{P}_{n}\psi(\hat{x}_{i}))^{R}+\frac{i}{2}\theta_{ij}(\hat{P}_{i}(\hat{x}^{R})_{k})\otimes (\hat{P}_{j}\psi^{R}(\hat{x}_{i}))
\end{eqnarray*}
Now substituting $(\hat{P}_{n}\psi)^{R}=R_{nm}(\hat{P}_{m}\psi^{R})$ and $\hat{P}_{i}(\hat{x}^{R})_{k}=-iR_{ki}$ in the above expression and taking the multiplication map eventually in (\ref{app}) yields, on further simplification, the desired result
\begin{eqnarray}
m[\bigtriangleup_{\theta}(R)(\hat{x}_{k}\otimes \psi(\hat{x}_{i}))]=\hat{\tilde{X}}^{(l){R}}_{k}\psi^{R}(\hat{x}_{i})
\end{eqnarray}
with
\begin{eqnarray}
\hat{\tilde{X}}^{(l){R}}_{i}=\hat{X}^{(l){R}}_{i}+\frac{1}{2}[R,\Theta]_{ij}\hat{P}_{j}
\end{eqnarray}



\begin{thebibliography}{99}
\bibitem{Doplicher} S.Doplicher, K.Fredenhagen and J.E.Roberts, Comm.Math.Phys. 172,187 (1995) 
\bibitem{Seiberg} N.Seiberg and E.Witten JHEP 09,032 (1999)
\bibitem{Hall} J.Bellissard, A.Van Elst and H Schulz-Baldes J.Math.Phys. 35, 5373 (1994)
\bibitem{Prodan} E.Prodan arxiv:1010.0695 [cond-mat]
\bibitem{Asch} P.Aschieri, J.Phys.Conf.Ser. 53 ,799 (2006)
\bibitem{Scholtz} F.G.Scholtz, B.Chakraborty, J.Govaerts and S.Vaidya J.Phys.A 40, 14581 (2007)
\bibitem{Basu} P.Basu, B.Chakraborty and F.G.Scholtz, J.Phys.A 44, 285204 (2011)
\bibitem{Chakra} B.Chakraborty, Z.Kuznetsova, F.Toppan J.Maths.Phys. 51. 112102 (2010)
\bibitem{Gouba} F.G.Scholtz, L.Gouba, A.Hafver, C.M.Rohwer J.Phys.A 42,175303 (2009)
\bibitem{Pinzul} A.P.Balachandran, A.Pinzul, B.A.Qureshi and S.Vaidya, Phys.Rev.D76, 105025 (2007); Phys. Rev. D77, 025021 (2008).
\bibitem{Banerjee} For a review see R.Banerjee, B.Chakraborty, S.Ghosh, P.Mukherjee and S.Samanta, Found. Phys. 39, 1297 (2009); A.P.Balachandran, A.Joseph and P.Padmanabhan Found.Phys.40,692 (2010)
\bibitem{Aschieri} "Lecture on Hopf Algebras, Quantum Groups and twists" P. Aschieri, e-print arxiv: hep-th/0703013
\bibitem{Castro} P.G.Castro, B.Chakraborty and F.Toppan, J.Math.Phys. 49, 082106 (2008)
\bibitem{Wess} J.Wess, Mathematical, Theoretical and Phenomenological Challenge Beyond the Standard model,The Workshop,2003.
\bibitem{Chaichian} M.Chaichian, P.P.Kulish, K.Nishijima and A.Turennu, Phys.Lett.B 604, 98(2004); M.Chaichian, P.Presnajder and A.Turennu, Phys.Rev.Lett. 94, 151602 (2005)
\end{thebibliography}
\end{document}